\documentclass{article}
\usepackage{graphicx,psfig,epsfig,rotating}
\def\title{\begin{center}\Large\bf}
\def\author(s){\vspace{0.3cm}\large\rm}
\def\text{\end{center}}

\begin{document}

\title
Solar coronal observations at high frequencies\\

\author(s)
A.\ C.\ Katsiyannis$^{\rm 1}$, M.\ Mathioudakis$^{\rm 1}$, 
K.\ J.\ H.\ Phillips$^{\rm 2}$, D.\ R.\ Williams$^{\rm 1}$,\\
F.\ P.\ Keenan$^{\rm 1}$\\
\vspace{0.1cm}
$^{\rm 1}${\it Department of Pure and Applied Physics, 
           Queen's University Belfast, Belfast, BT7 1NN, UK}\\
$^{\rm 2}${\it Space Science \& Technology Department, Rutherford Appleton Laboratory, Chilton, Didcot, Oxon. OX11 0QX, UK}\\
\text

\vspace{0.3cm}

\large

\section*{Abstract}
The Solar Eclipse Coronal Imaging System (SECIS) is a simple and
extremely fast, high-resolution imaging instrument designed for
studies of the solar corona. Light from the corona (during, for
example, a total solar eclipse) is reflected off a heliostat and
passes via a Schmidt-Cassegrain telescope and beam splitter to two CCD
cameras capable of imaging at 60 frames a second. The cameras are
attached via SCSI connections to a purpose-built PC that acts as the
data acquisition and storage system. Each optical channel has a
different filter allowing observations of the same events in both
white light and in the green line (Fe XIV at 5303 {\AA}). Wavelet
analysis of the stabilized images has revealed high frequency
oscillations which may make a significant contribution on the coronal
heating process. In this presentation we give an outline of the
instrument and its future development.

\section{Introduction}

For the last 50 years one of the most important unexplained problems
in solar physics is to explain the exact heating mechanism for the
corona. At a temperature of $\sim$ 10$^{6}$K the solar atmosphere is
much hotter than the photosphere or chromosphere, ruling out the
possibility of radiative heating. It is widely accepted that magnetic
fields must be the reason for the high temperature, but there is
disagreement about the exact mechanism. Parker (1988 and references
therein) supported the idea of numerous small-scale magnetic
reconnections (nanoflares) releasing enough energy to heat the corona,
while Hollweg (1981) and others favoured dumping of energy from
magnetohydrodynamic (MHD) waves. Both theories are supported by
observational evidence.\\

Our project concentrates on the detection of MHD waves that can
dissipate enough energy in their environment and heat the
corona. Porter et al.\ (1994a \& 1994b) produced theoretical
simulations of such waves under a range of conditions and found that
only high frequency ($\ge$ 0.5 Hz) oscillations are capable of heating
the coronal loops. Currently none of the spaced-based telescopes has
instruments capable of detecting such fast events.  Therefore a
ground-based instrument was needed as a short term solution which
could either observe the solar corona during a total eclipse or with a
coronagraph. The first such instrument was developed by Pasachoff \&
Landman (1984) who reported some evidence of periodicity in the 0.5-2
Hz range during the total solar eclipse of February 1980. More
evidence of this type of oscillation were found again during the June
1983 eclipse by Pasachoff \& Ladd (1987) and Pasachoff (1997). Rusin
\& Minarovjech (1994) and Singh et al. (1997) have also reported
short-period waves while Koutchmy et al. (1983) used the coronagraph
at the National Solar Observatory/Sacramento Peak to produce evidence
of oscillations with periods of 43s, 80s and 300s.\\

In this paper we will describe an instrument we have developed in
conjunction with the Astronomical Institute of the University of
Wroc{\l}aw, Poland, called the Solar Eclipse Coronal Imaging System
(SECIS), which performs rapid observations of coronal loops. Some
results from the August 1999 eclipse are also included (also see
Phillips et al.\ 2000; Williams et al.\ 2001, 2002).\\

\section{The SECIS instrument}

Figure 1 contains a schematic diagram of the SECIS optical path as
published by Williams et. al.\ (2001). The primary mirror of the
Schmidt-Cassegrain telescope has a diameter of 20cm, the beam splitter
allows 90\% of the light to pass through and reflects the remaining
10\%. The green filter in front of camera 1 has a bandwidth centred
around the Fe {\sc xiv} emission line at 5303 \AA, and the measured
full width half-maximum (FWHM) of this filter is 4 \AA. Camera 2 has
no filter, allowing simultaneous white light observations.\\

\begin{figure}
\centering
\mbox{\begin{turn}{-90}\epsfclipon\epsfxsize=8cm\epsfbox[0 0 500 750]
     {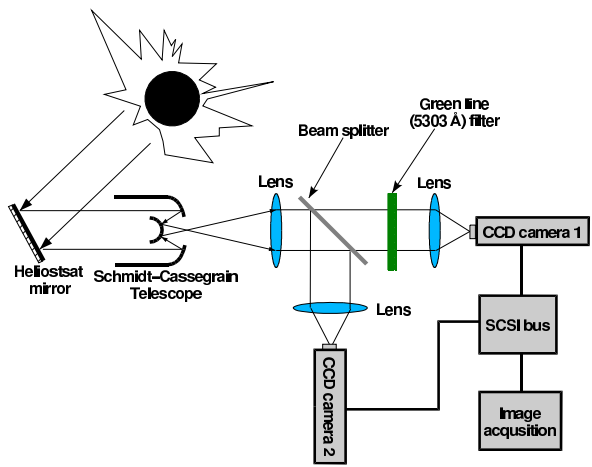}\end{turn}}
\caption{Schematic diagram of the SECIS instrument as published by 
Williams et al.\ (2001)}
\label{fig1}
\end{figure}

Both cameras are connected via a Personal Computer (PC) that acts as
the data acquisition system. As each camera is capable of recording up
to 60 frames per second (fps) with a 512 $\times$ 512 resolution and a
pixel depth of 12 bits, a high data bandwidth was needed. To solve
this problem we attached a Redundant Array of Inexpensive Disks (RAID)
of 4 Small Computer System Interface (SCSI) disks each of 9
GByte capacity. During data acquisition, each image from each camera is
split in two, and while the first two disks record the first half of
the images produced by the two cameras, the other two disks place
their heads to pre-allocated positions for the writing of the rest of
the images. Once the first two disks finish, the other two start
recording the second half while the first two disks position
themselves for the next two images.\\

\section{Wavelet analysis}

The detection of oscillations requires the analysis of time series
observations of individual pixels. Although Fourier analysis appears
to be the most popular frequency analysis method, there is a new
tendency toward the use of wavelets (for example Mallat 1998, Ireland
et al.\ 1999, Gallagher et al.\ 1999). The main reason is that
although this analysis is computationally more intensive, it helps to
isolate oscillations localized in both time and frequency. The authors
believe that this is the best way to detect transient oscillations and
rapid changes, while comparing with the quality of data at the same
time.\\

\begin{figure}
\centering
\mbox{\begin{turn}{90}\epsfxsize=12cm\epsfbox[0 0 650 800]
     {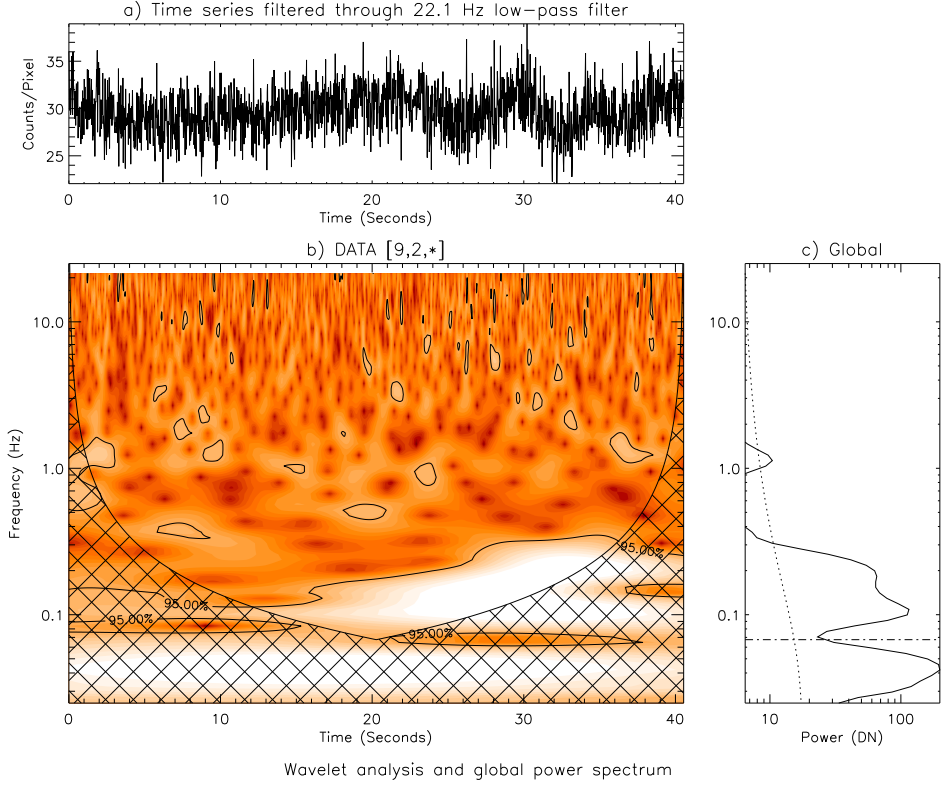}\end{turn}}
\caption{Wavelet analysis of a pixel located near the peak of a 
         coronal loop during the August 1999 total solar eclipse.
         See text for details.}
\label{fig2}
\end{figure}

The wavelet used in our transform is the Morlet wavelet given by the
formula:

\begin{equation}
\psi(\eta)=\pi^{-1/4}\exp(i\omega_0\eta)\exp(\frac{-\eta^{2}}{2})
\label{morlet}
\end{equation}

where $\eta$ is the dimensionless time parameter, $\omega_0$ is the
dimensionless frequency parameter and $\pi^{-1/4}$ is a normalization
term (Torrence \& Compo 1998).\\

Williams et al.\ (2001, 2002) have already detected a number of
oscillations during the August 1999 eclipse using SECIS observations
and wavelet analysis. The waves have periods around 0.16 Hz providing
us with observational evidence that coronal heating via short-period
MHD waves is feasible. Figure 2 demonstrates yet another detection of
a coronal oscillation at a nearby frequency using the same data as
Williams et al.\ (2001, 2002). It is located in the apex peak of
another loop in the same active region (NOAA AR 8561).\\

In Figure 2a a plot of the time series analyzed is included while at
the right-hand side (Figure 2c) is the global wavelet spectrum,
analogous to a Fourier power spectrum. In Figure 2b (the wavelet power
transform), the lightly shaded area indicates a good match between the
wave given by the equation (1) for a given frequency (specified by the
y-axis) for a given point at the time sequence (specified by the
x-axis). The contoured line surrounding the white area indicates areas
with more than 95\% confidence. The cross-hatched area is polluted by
edge effects and should be ignored.\\

The authors would like to emphasize that this is not the only
detection of high frequency periodicities in this particular coronal
loop. There are another three oscillations found in pixels surrounding
the one from which Figure 2 was produced, and we believe this is
strong evidence in favour of the credibility of the presented
detection. It indicates that there was an actual wave along the loop
and became visible only at the peak because of surrounding emission
from other coronal loops.\\

\section*{Acknowledgments}

ACK acknowledges funding from the Leverhulme Trust via grant
F00203/A. DRW acknowledge studentships funded by the
Department of Higher \& Further Education, Training \&
Employment. 

\section*{References}

\noindent Gallagher P.T., Phillips K.J.H, Harra-Murnion L.K., Baudin F., Keenan F.P.; 1999; A\&A; 348; 251.

\noindent Hollweg J.; 1981; SoPh; 70; 25.

\noindent Ireland J., Walsh R.W., Harrison R.A., Priest E.R.; 1999; A\&A; 347; 255.

\noindent Koutchmy S., Belmahdi M., Coulter R.L., Demoulin P., Gaizauskas V., MacQueen R.M., Monnet G., Mouette J., Noens J.C., November L.J.; 1983; A\&A; 281; 249.

\noindent Mallat S.; 1998; A Wavelet tour of signal processing; Academic Press; London.

\noindent Parker E.N.; 1988; ApJ; 281; 249.

\noindent Pasachoff J.M.; 1997; in {\rm Theoretical and Observational Problems Realted to Solar Eclipses}; eds Mouradian \& Stavinschi; Kluwer; 181.

\noindent Pasachoff J.M., Land E.F.; 1987; SoPh; 109; 365.

\noindent Pasachoff J.M., Landman D.A.; 1984; SoPh; 90; 325.

\noindent Phillips K.J.H., Read P.D., Gallagher P.T., Keenan F.P., Rudawy P., Rompolt A., Berlicki A., Buczylko A., Diego F., Barnsley R., Smartt R.N., Pasachoff J.M., Babcock B.A.; 2000; SoPh; 193; 259.

\noindent Porter L.J., Klimchuk J.A., Sturrock P.A.; 1994a; ApJ; 435; 482.

\noindent Porter L.J., Klimchuk J.A., Sturrock P.A.; 1994b; ApJ; 435; 502.

\noindent Rusin V., Minarovjech M.; 1994; IAU Symposium no.\ 144; 487.

\noindent Singh J., Cowsik R., Raveendran A. V., Bagare S. P., Saxena A. K., Sundararaman K., Krishan V., Naidu N., Samson J. P. A., Gabriel F.; 1997; SoPh; 170; 235.

\noindent Torrence C., Compo G.P.; 1998; Bull.\ Amer.\ Meteor.\ Soc.; 79; 61.

\noindent Williams D.R., Phillips K.J.H., Rudawy P., Mathioudakis M., Gallagher P.T., O'Shea E., Keenan F.P., Read P., Rompolt B.; 2001; MNRAS; 326; 428.

\noindent Williams D.R., Mathioudakis M., Gallagher P.T., Phillips K.J.H., McAteer R.T.J., Keenan F.P., Katsiyannis A.C.; 2002; MNRAS; submitted.

\end{document}